\documentclass[twocolumn,showpacs,prb]{revtex4}
\usepackage{graphics,german,amsmath,amssymb}

\usepackage{graphicx}
\usepackage{color}
\usepackage{epstopdf}
\usepackage{bm}

\newcommand{\figref}[1]{Fig.\,\protect\ref{#1}}

\begin{document}

\title{Magnetic properties of vanadium-oxide nanotubes probed by static magnetization and $^{51}$V~NMR}

\author{E. Vavilova$^{1,2}$}
\author{I. Hellmann$^1$}
\author{V. Kataev$^{1,2}$}
\author{C. T"aschner$^1$}
\author{B. B"uchner$^1$}
\author{R. Klingeler$^1$}

\affiliation{$^{1}$Leibniz-Institute for Solid State and Materials Research IFW Dresden, P.O. Box 270116, D-01171
Dresden, Germany}

\affiliation{$^{2}$ Kazan Physical Technical Institute of the Russian Academy of Sciences, 420029 Kazan, Russia}

\date{\today}

\begin{abstract}
Measurements of the static magnetic susceptibility and of the nuclear magnetic resonance of multiwalled  vanadium-oxide nanotubes are reported. In
this nanoscale magnet the structural low-dimensionality and mixed valency of vanadium ions yield a complex temperature dependence of the static
magnetization and the nuclear relaxation rates. Analysis of the different contributions to the magnetism allows to identify individual interlayer
magnetic sites as well as strongly antiferromagnetically coupled vanadium spins ($S\,=\,1/2$) in the double layers of the nanotube's wall. In
particular, the data give strong indications that in the structurally well-defined vanadium-spin chains in the walls, owing to an inhomogeneous
charge distribution, antiferromagnetic dimers and trimers occur. Altogether, about 30\,\% of the vanadium ions are coupled in dimers, exhibiting a
spin gap of the order of 700\,K, the other $\sim\,30$\,\% comprise individual spins and trimers, whereas the remaining $\sim\,40$\,\% are
nonmagnetic.
\end{abstract}

\pacs{75.75.+a, 75.40.Cx, 76.60.-k}

\maketitle

\section{Introduction}

{Nanoscale magnetic materials are the subject of a currently increasing interest which is related both to the novel} {fundamental aspects of magnetic
structure and dynamics on the nanometer-scale, {and} promising future technological applications, such as magnetic data storage {elements}, sensors
or spin electronic devices {(for a review see, e.g., Ref.~\onlinecite{Himpsel98})}. One novel approach {to engineer} single, i.e. well separated and
individually addressable nanoscale magnets, is to utilize self-organization of nanostructures, {such as} carbon fullerenes, carbon nanotubes,
{silicon-carbon}-nanorods {and others}. {However, owing to a weak magnetic exchange between unpaired spins} these materials {often} exhibit only a
very small magnetic moment so that external magnetic fields barely interact with these nano\-structures.} {One possible way to overcome this problem
is to prepare self-organized nanoscale materials based on transition-metal (TM) oxide low-dimensional (low-D) structural units where complex strong
electronic correlations in the spin- and charge sectors are expected. Indeed,} a variety of fascinating phenomena related to spin- and charge
correlations in {low-D TM} oxides were found in the past years, ranging from high-temperature superconductivity in 2D-cuprates, a spin-Peierls
transition or a Haldane spin-gap behavior in 1D-spin chain compounds, to superconductivity in 1D-spin ladder materials (for recent reviews see e.g.
Ref.~\onlinecite{Imada98,Dagotto99,Tokura00,Orenstein00}).

\begin{figure}
    \includegraphics[width=0.9\columnwidth]{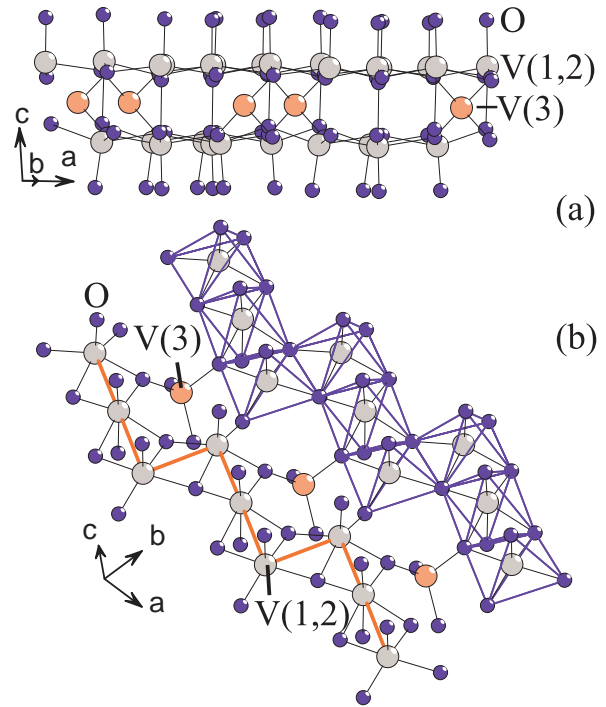}
    \caption{(Color online) Structure of the VO$_x$-NT wall: (a) Double layer of octahedrally coordinated V(1,2) ions. Tetrahedrally coordinated V(3)
    ions are located between the two layers. (b) Zig-zag V(1,2)-chains in a layer.}
\label{StructureVOx}
\end{figure}

Recently a new class  of nanoscale spin magnets, {mixed valence} vanadium oxide multiwall nanotubes (VO$_x$-NTs), has been
introduced.\cite{Krumeich99} The structure of the VO$_x$-NT walls \cite{Worle02} is closely related to that of barium vanadium oxide bronze
BaV$_7$O$_{16}\,\times\,n\,$H$_2$O (Fig.~\ref{StructureVOx}).\cite{Wang98} One finds there a V-metal ion in distorted octahedral and tetrahedral
coordination, respectively. V-octahedra are coupled in edge-sharing zig-zag VO-chains. Arrays of VO-chains form a double layer with tetrahedral
V-sites lying between the two layers. Mixed valence vanadium oxides, VO$_x$ ($1.0\,<\,x\,<\,2.5$), are themselves very rich in electronic and
magnetic properties owing to an intimate interplay between spin, orbital and charge degrees of freedom.\cite{Imada98} When assembled in nanotubes
(NT) using dodecylamine, C$_{12}$H$_{25}$NH$_2$, as a structure-directing surfactant molecule, VO$_x$-NTs show up diverse properties ranging from
spin frustration and semiconductivity to ferromagnetism by doping with either electrons or holes.\cite{KrusinElbaum04} Such a rich behavior makes
this material very interesting from the viewpoint of fundamental research as well as regarding possible future applications {mentioned above}. {In
particular, {its magnetic properties can be widely tuned by} intercalation or electrochemical doping (e.g. by lithium or iodine), thereby offering
the possibility {to realize novel nanoscale magneto-electronic devices}.} {A detailed analysis of the magnetic properties is, {therefore, important}
in order to correlate them with structure and morphology of VO$_x$-NTs  for tailoring the magnetic properties {of this nanostructured material}.}

In order to obtain a deeper insight into {the magnetism} of VO$_x$-NTs we have measured magnetization and nuclear magnetic resonance (NMR) of
pristine, i.e. Li-undoped, samples. The static magnetic data give evidence for the occurrence of magnetically nonequivalent vanadium sites in the
structure. These sites can be presumably attributed to V$^{4+}$ ($d^1$, $S\,=\,1/2$) ions in the octahedral and tetrahedral oxygen coordination,
respectively. The spins in the octahedral sites (chain sites), are strongly antiferromagnetically correlated. One part of them is coupled in dimers
and exhibits a spin-gap behavior. The other part forms trimers. They, together with much weaker correlated tetrahedral magnetic sites, dominate the
low-temperature static magnetic response. NMR measurements on $^{51}$V nuclei reveal relaxation channels with two distinctly different relaxation
rates.  Though the longitudinal relaxation rate $T_1^{-1}$ exhibits a strong temperature dependence the manifestation of a spin gap is obscured due
to the presence of an appreciable amount of magnetic sites at low temperatures. We discuss possible mechanisms of vanadium nuclear spin relaxation
and their interplay with the static magnetic properties. In particular, both NMR and the static susceptibility data give strong indications that
owing to the pronounced structural low-dimensionality and mixed valency of VO$_x$-NTs different spin arrangements, namely individual spins, spin
dimers and trimers, coexist in comparable amounts in this complex material.

\begin{figure}
    \includegraphics[width=\columnwidth]{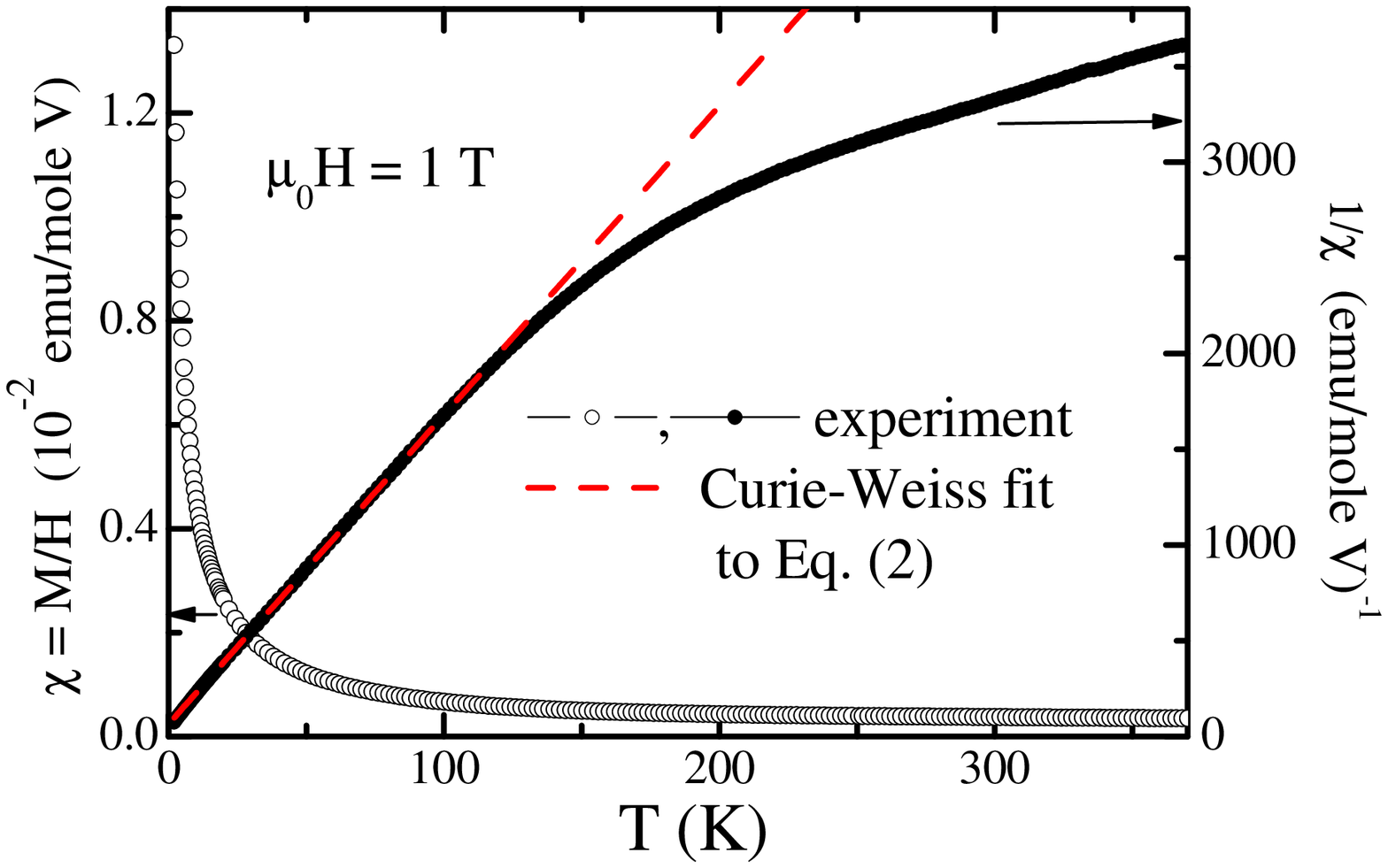}
    \caption{(Color online) Temperature dependence of the static susceptibility $\chi(T)=M(T)/H$ in an external magnetic field of
    $\mu_0H=1$\,T (left ordinate) and of $1/\chi(T)$ (right ordinate). The dashed line is a
    Curie-Weiss fit of the low temperature data to Eq.~(\ref{chiCW}). (For details see the text.)}
\label{chiT}
\end{figure}

\section{Experimental results and discussion}

\subsection{Samples and Experimental}

Samples of VO$_x$-NTs were synthesized using a hydrothermal procedure described in Ref.~\onlinecite{Krumeich99,KrusinElbaum04}. The high quality of
the samples has been verified by a number of techniques, such as high-resolution and analytical transmission electron microscopy (TEM), electron
diffraction, x-ray photoemission spectroscopy and electron energy-loss spectroscopy (EELS)(for details see Ref.~\onlinecite{Liu05}). {In this
preparation method the nanotubes occur in two topological modifications, i.e. concentric shell tubes with closed edges of the VO$_x$ sheets and
scrolled tubes with open edges, with the latter topology being a predominant one. \cite{Krumeich99} As one can see in Fig.~1 of
Ref.~\onlinecite{Liu05}, however, the tube walls in our samples consist of a large number ($\sim\,20$) of VO$_x$ layers. Therefore one expects that
such differences in the microscopic topology, yielding absence or presence of open vanadium edge sites plausibly different from those in the bulk of
the wall, should not affect appreciably the bulk magnetic properties. }

Here, static magnetization measurements have been performed in the temperature range 2\,-\,500\,K with a SQUID magnetometer from Quantum Design in
magnetic fields up to 5~T and with a home-made vibrating sample magnetometer (VSM) in fields up to 15~T. Upon heating, however, we observed
indication that, above $\sim$370\,K, the material degenerates so that we restrict the data presented here to temperatures below 370\,K. {Presumably
this sample degradation is due to the changes of the oxygen stoichiometry. We note, however, that on the time scale of two months repeated
magnetization measurements of the samples kept in air yielded identical results which assures the sample stability if overheating is avoided.}

The $^{51}$V-NMR experiments were carried out {with a Tecmag pulse solid-state NMR spectrometer} in a field of 7.05~T in a temperature range
15\,-\,285\,K.

\subsection{Static magnetization}
\label{statmag}

\begin{figure}
    \includegraphics[width=0.8\columnwidth]{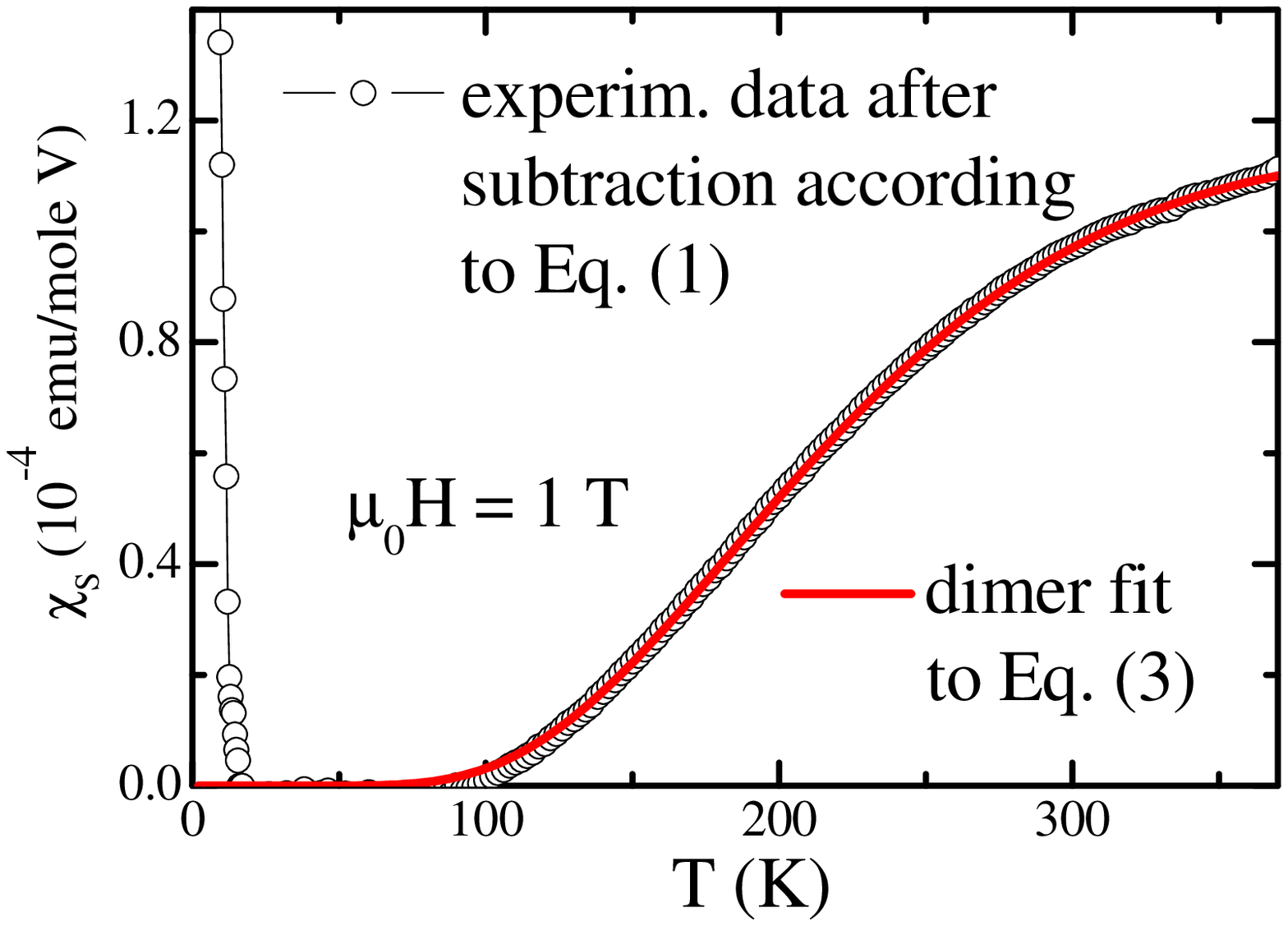}
    \caption{(Color online) High temperature contribution $\chi_s$ to the total susceptibility, Eq.~(\ref{chidiff}).
    A fit to a dimer model, Eq.~(\ref{chidimer}), is shown by a solid line (red color online) which is practically undistinguishable from the data.
    (For details see the text)} \label{chiS}
\end{figure}

\textit{Experiment.} The temperature dependence of the static susceptibility $\chi(T)\,=\,M/H$ obtained in a magnetic field of $\mu_0H\,=\,1$\,T is
shown in \figref{chiT}. These data are similar to the findings in Ref.~\onlinecite{KrusinElbaum04}. One can distinguish between a low-temperature
regime below $\sim\,120$\,K characterized by a Curie-like behavior, which is reflected by a linear temperature dependence of $1/\chi(T)$ (cf.
\figref{chiT}). At higher temperatures, however, there are strong deviations from the linear behavior. The latter can be attributed to an additional
contribution to the static susceptibility $\chi_s$ arising at higher temperatures. In this case $\chi_s$ can be obtained by subtracting from the raw
data the Curie-Weiss term $\chi_{CW}$ and a small temperature independent contribution $\chi_0$ owing to the Van-Vleck paramagnetism of the V ions:
\begin{eqnarray}
\chi_s\,=\,\chi(T)\,-\,\chi_{CW}\,-\,\chi_0,\label{chidiff}\\
\chi_{CW}\,=\,C/(T\,+\,\Theta).\label{chiCW}
\end{eqnarray}
Here, $C\,=\,N_{CW}\,N_A\,\mu_B^2/k_B$ is the Curie constant, assuming the spin $S\,=\,1/2$ and the $g$ factor $g\,=\,2$, and $\Theta$ is the
Curie-Weiss temperature. The best fit of the low-temperature data yields the concentration of V$^{4+}$ spins contributing to the Curie-Weiss
susceptibility $N_{CW}\,=\,17$\,\% of all V ions, $\Theta\,=\,4$\,K and $\chi_0\,=\,7\,\cdot\,10^{-5}$\, emu/mole. The additional contribution
$\chi_s$ is shown in \figref{chiS}.

Consistently with the results of Ref.~\onlinecite{KrusinElbaum04}, $\chi_s$ can be explained with a model of noninteracting antiferromagnetic dimers:
\begin{equation}
\chi_{dimer}\,=\,(N_dN_A\mu_B^2/k_B)/\left(T[3 + \exp(\Delta/k_BT)]\right).\label{chidimer}
\end{equation}
Fitting of the data in \figref{chiS} to Eq.~(\ref{chidimer}) implies that at $N_d\,=\,28$\,\% of all V sites there are spins $S=1/2$ which form
antiferromagnetic dimers. The dimer gap amounts to $\Delta\,=\,710$\,K. With these parameters the fit yields $\chi_s\,=\,\chi_{dimer}$ with a good
accuracy.

\begin{figure}
    \includegraphics[width=0.8\columnwidth]{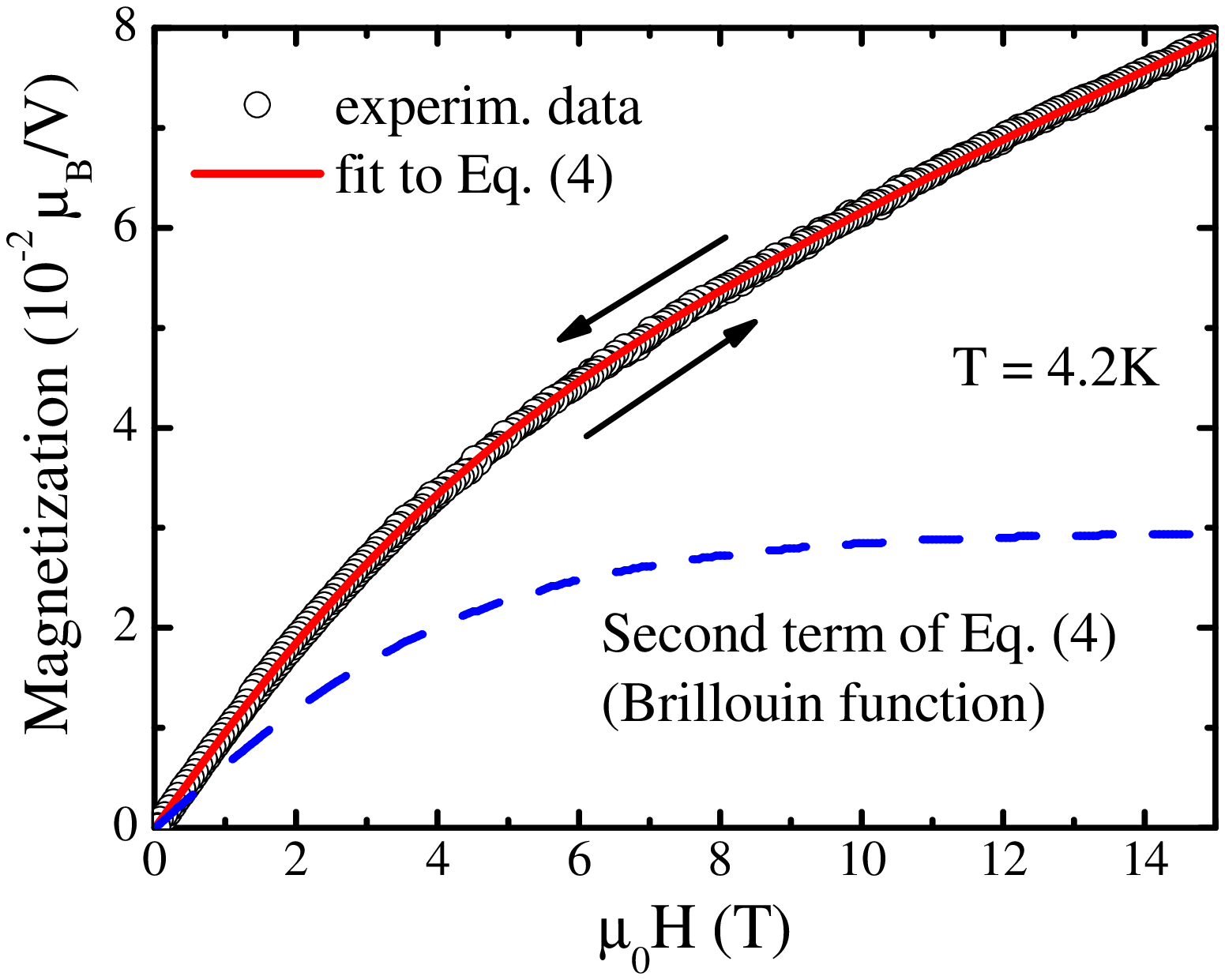}
    \caption{(Color online) Magnetic field dependence of magnetization at $T\,=\,4.2$\,K measured in ascending and descending fields.
    Note the absence of hysteresis. A fit to Eq.~(\ref{Mfit}) is shown by a solid line (red color online) which is practically undistinguishable
    from the data. Contribution owing to quasi-free spins described by the Brillouin function is shown by a dashed line. (For details see the text)}
\label{MH}
\end{figure}

Hence, the susceptibility data suggest that in a large temperature range the static magnetic response of VO$_x$-NTs can be very well described by a
Curie-Weiss contribution $\chi_{CW}$, the Van-Vleck paramagnetism $\chi_0$ and the presence of antiferromagnetic dimers. However, below $\sim\,15$\,K
the experimental data deviate from this description, (\figref{chiS}). Here, $\chi_{dimer}$ is practically zero owing to a large value of $\Delta$.
Thus the strong upturn of $\chi_s$ below 15\,K implies the occurrence of another contribution to the static susceptibility not accounted for in
Eqs.~(\ref{chidiff}) and (\ref{chidimer}). In order to address the low temperature magnetism, the field dependence of the magnetization $M(H)$ up to
15\,T, at $T\,=\,4.2$\,K, has been measured. The data presented in \figref{MH} display a nearly linear field dependence of $M$ at high magnetic
fields and a pronounced curvature at lower fields without measurable hysteresis. This $M(H)$ curve can be very well fitted (\figref{MH}) as a sum of
two contributions
\begin{equation}
M(H)\,=\,\chi_{lin} H \,+\, N_BN_A\mu_BB_s(x).\label{Mfit}
\end{equation}

The first part of Eq.~(\ref{Mfit}) describes an appreciable linear contribution to $M$ determined by a field-independent susceptibility
$\chi_{lin}\,=\,1.8\,\cdot\,10^{-3}$\, emu/mole. The second part accounts for a nonlinearity of $M(H)$ due to the alignment of the spins by a
magnetic field at low temperatures. Here $B_s(x)$ is the Brillouin function with $x\,=\,H\,+\,\lambda M$, and $\lambda$ being the mean-field
parameter. Assuming the $g$ factor $g\,=\,2$ and the spin $S\,=\,1/2$ one obtains from the fit a small mean-field parameter $\lambda\,=\,70$\,
mole/emu and the concentration of such polarisable, i.e. quasi-free, spins $N_B\,=\,3\%$ of all V sites at $T\,=\,4.2$\,K.

\textit{Discussion.} We will start the discussion of the magnetization data by considering the response at $T>15$\,K, where the susceptibility can be
described well in terms of non-interacting dimers and the Curie-Weiss law. Different contributions to the magnetic susceptibility can be related to
the occurrence of nonequivalent vanadium sites in the crystal structure of VO$_x$-NTs.\cite{KrusinElbaum04} The octahedrally coordinated V(1) and
V(2) sites are found in the double layers (Fig.~\ref{StructureVOx}). Tetrahedrally coordinated V(3) sites occur between the layers. The bond valence
sums calculated for a closely related material BaV$_7$O$_{16}\,\times\,n\,$H$_2$O suggest that vanadium in the V(3) position has a valency close to
4+ ($d^1$, $S\,=\,1/2$), whereas vanadium in the positions V(1) and V(2) may be 4+ as well as 5+ ($d^0$, $S\,=\,0$).\cite{Worle02,Wang98} The
mixed-valency in VO$_x$-NTs is confirmed by recent EELS measurements which yield 60\% of V$^{4+}$ and 40\% of V$^{5+}$ in our samples.\cite{Liu05}
Thus the concentration of magnetic V sites with $S=1/2$ amounts to 60\%. This estimate is consistent with our NMR data (see Section~\ref{NMR} below)
but, however, at first glance numerically contradicts the static magnetization data. Indeed, the number of magnetically active sites contributing to
the Curie-Weiss susceptibility and to the dimer susceptibility makes up $N_{CW}+N_d\,=\,17$\,\%+$28$\,\%=45\% only. The number $N_{CW}$, however,
does not necessarily count only single $S=1/2$ V sites, which we will discuss in the following.

It is reasonable to assume that owing to magnetic frustration the V(3) spins between the layers are interacting very weakly and that all magnetic
V(3) sites are therefore contributing to the Curie-like behavior of the static susceptibility.\cite{KrusinElbaum04} The amount of V(3) sites in the
unit cell is $1/7\,\approx\,14$\,\%. A larger number of weakly interacting Curie-like 1/2-spins $N_{CW}\,=\,17$\,\% estimated from the analysis of
the susceptibility (Fig.~\ref{chiT}) suggests that quasi-free spins are not confined to the interlayer V(3) sites only. The mixed valency of vanadium
in VO$_x$-NTs results in a heterogeneous charge distribution in the zig-zag chains comprising the walls (Fig.~\ref{StructureVOx}). Assuming that most
of tetrahedral V(3) sites are magnetic ($\lesssim\,14$\,\%) and recalling that the EELS data yield 40\% of nonmagnetic V$^{5+}$ ions in the structure
which thus should mostly reside in the chains, one finds almost equal number of spin- ($S\,=\,1/2$) and spinless ($S\,=\,0$) sites in the chains.
Thus, about a half of the sites in the chain are magnetic. Recalling our estimate of the concentration of spins in dimers $N_d\,=\,28$\,\% which all
should reside in the chains, we obtain that those spins constitute there almost 2/3 of magnetic chain sites. Hence, there is still a significant
number of spins in the chains amounting to $\sim\,18\%$ of all V sites in the unit cell, which are not coupled in dimers. If they were individual
quasi-free spins, they would be expected to contribute to the Curie-Weiss susceptibility, similar to the V(3) interlayer sites. This would yield the
total concentration of Curie-like spins $18\%\,+\,14\%\,=\,32\%$, which is almost twice as large as the experimentally determined value
$N_{CW}\,=\,17$\,\%. Therefore, this apparent discrepancy suggests that the remaining spins in the chain which are not involved in antiferromagnetic
dimers are not independent but might form longer spin-chain fragments. Indeed, it has been recently shown both theoretically and experimentally that
the ground state configuration of a strongly hole-doped antiferromagnetic Heisenberg chain is composed of spin chain fragments of different length,
i.e. of spin dimeres, trimers and monomers (individual spins) etc.\cite{Gelle04,Klingeler06} Remarkably, for small external fields, trimers respond
similarly to free spins at low temperatures (see e.g. Ref.~\onlinecite{Klingeler05b}). Thus $N_{CW}\,=\,17$\,\% of Curie-like spins estimated from
our susceptibility data may comprise magnetic interlayer V(3) sites, i.e. quasi-free spins, as well as trimers in the V chain.

\begin{figure}
    \includegraphics[width=0.7\columnwidth]{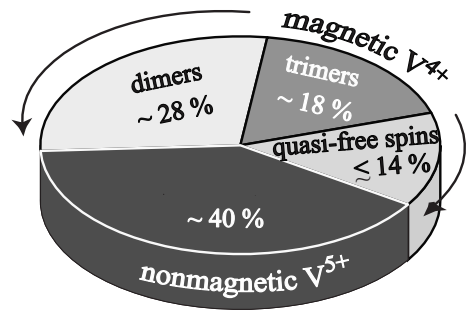}
    \caption{(Color online) Fractional weight of different spin species in VO$_x$-NTs.  (For details see the text)}
\label{Diagram}
\end{figure}

In case of antiferromagnetic trimers, the number of magnetically active spins involved in the Curie-like response is three times larger than for
single spins, i.e. for a number of spins $N_t$ arranged in trimers only $N_t/3$ will be counted in $N_{CW}$. Assuming that the remaining spins in the
chains which are not involved in dimers, all form trimers, i.e. $N_t\,=\,18$\,\%, one obtains a correct estimate of the total concentration of
magnetic $S\,=\,1/2$ sites in the sample $(N_{CW}\,-\,N_t/3)\,+\,N_t\,+\,N_d\,=\,57$\,\%, which is in a good agreement with the EELS data. To
summarize the above discussion the approximate fractional weights of different spin species are depicted in the diagram in \figref{Diagram}.

The above data analysis hence strongly suggests the occurrence in the V chains of spin-chain fragments of different lengths, i.e. individual spins,
antiferromagnetic dimers and trimers. One might speculate if even longer fragments occur. The presence of chain fragments with an even number of
sites, e.g. of quadrumers, however, can be ruled out by the experimental data: For quadrumers, one would expect the gap
$\Delta_q\,=\,0.659\,\cdot\,\Delta\,\approx\,470$\,K, \cite{Klingeler05b} which is not observed in our data. We also note, that the singlet-triplet
excitation in the trimer occurs at $\Delta_t\,=\,1.5\,\cdot\,\Delta$ which can not be investigated since the sample thermally degenerates at 370\,K.

Finally, we discuss the magnetization data at a low temperature $T\,=\,4.2$\,K (\figref{MH}) which reveals a substantially different magnetic
response as compared to the results at $T\,>\,15$\,K. In particular, the concentration of quasi-free spins, i.e. monomers (individual spins) or
trimers, is strongly reduced down to 3\%. This clearly indicates that most of the monomers/trimers are coupled antiferromagnetically at this
temperature. This scenario is corroborated by the observation of a significant linear contribution $\chi_{lin} H$ to the $M$ vs. $H$ curve. The fact
that $\chi_{lin}$ is constant in the whole field range implies that the antiferromagnetic coupling of the monomers/trimers is larger than the energy
$\mu_BH$ of the applied field of 15\,T. This scenario of antiferromagnetic coupling, however, does not straightforwardly explain the
\textit{additional} contribution to $\chi(T)$ below $\sim\,15$\,K (cf. \figref{chiS}). One may speculate that anisotropic coupling mechanisms, like
e.g. the Dzyaloshinskii-Moriya exchange, have to be considered. This requires, however, a detailed symmetry analysis of the exchange paths between
vanadium spins in VO$_x$-NTs.

\subsection{Nuclear Magnetic Resonance}
\label{NMR}

\begin{figure}
    \includegraphics[width=0.6\columnwidth,angle=-90,clip]{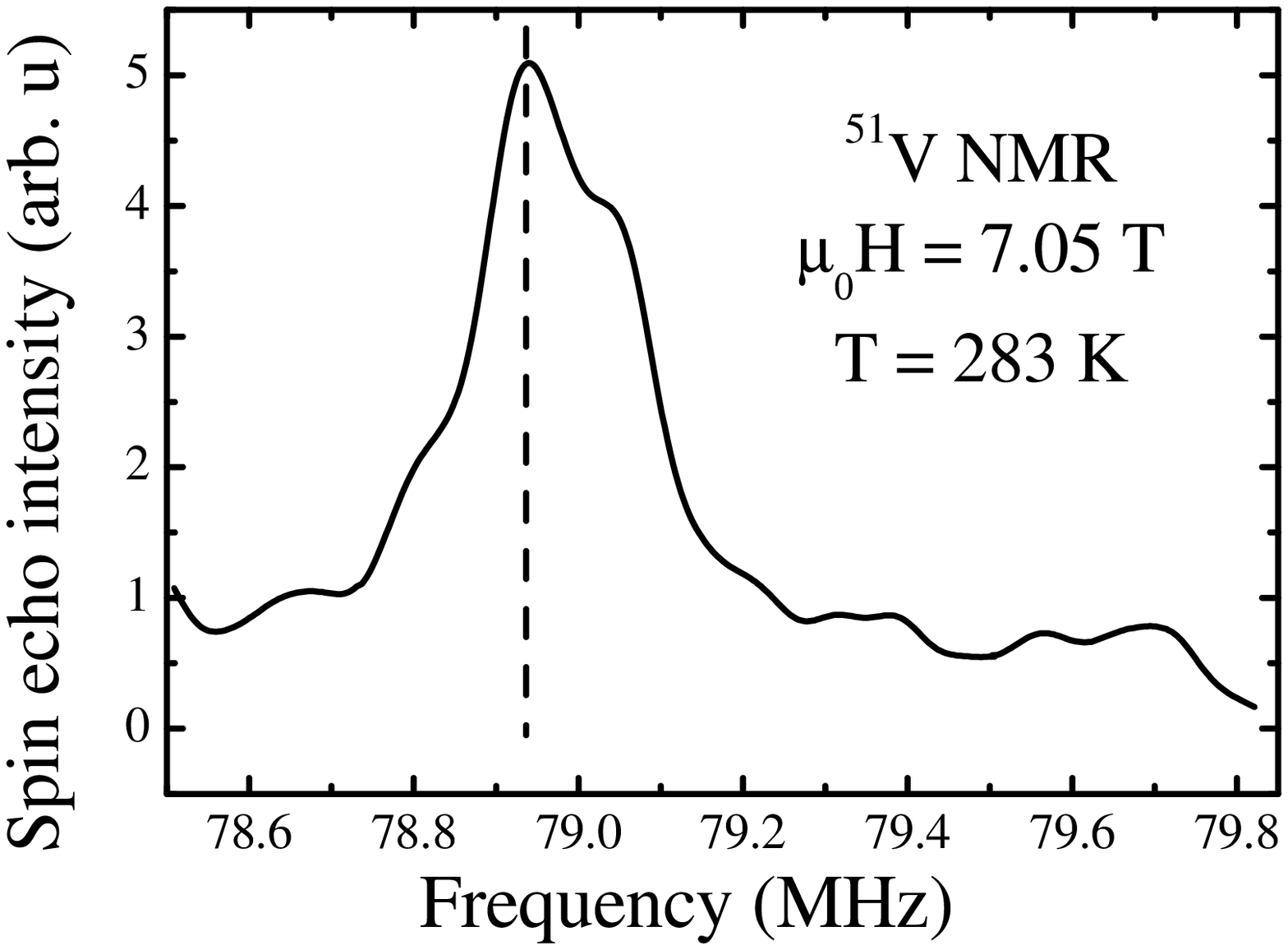}
    \caption{ $^{51}$V-NMR spectrum of a VO$_x$-NT sample at $T\,=\,285$\,K. The dashed line indicate the frequency
    of the maximum intensity at which the relaxation measurements have been carried out. }
\label{spectrum}
\end{figure}

\begin{figure}
    \includegraphics[width=\columnwidth,clip]{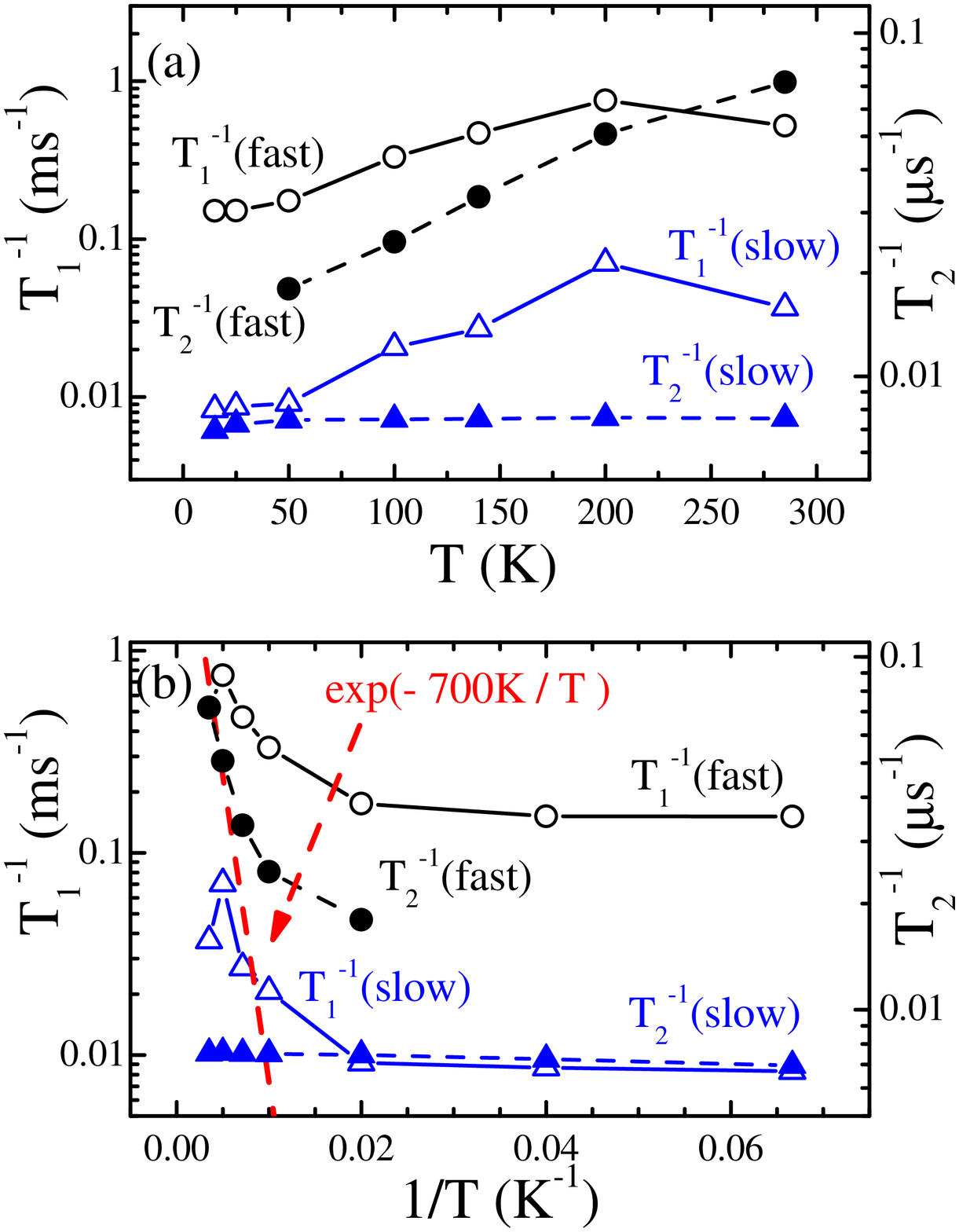}
    \caption{(Color online) ''Fast'' and ''slow'' contributions to the NMR relaxation rates $T_1^{-1}$ and $T_2^{-1}$:
    (a) as a function of temperature; (b) as a function of inverse temperature. Dashed line denotes the expected behavior
    of $T_1^{-1}$ {according to Eq.~(\ref{spingap})} due to the opening of the spin gap $\Delta\,\sim\,700$\,K.}
\label{rates}
\end{figure}

\begin{figure}
    \includegraphics[width=0.6\columnwidth,angle=-90,clip]{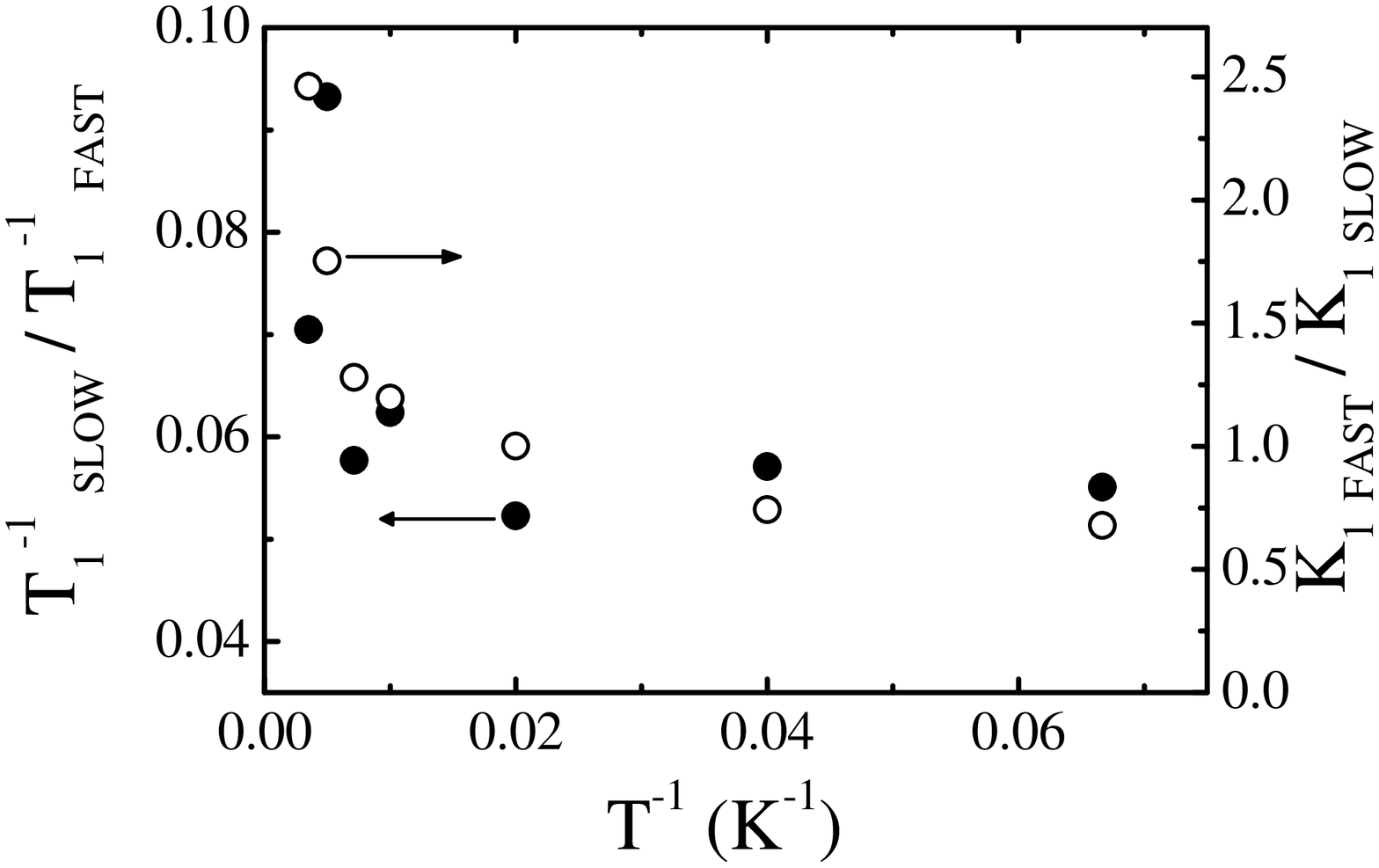}
    \caption{ Dependence on the inverse temperature of the ratio of the ''slow'' and ''fast'' contributions to the
    NMR relaxation rate $T_1^{-1}$ an the ratio of the weights of these contributions $k_{fast}\,:\,k_{slow}$.}
\label{ratio}
\end{figure}

\textit{Experiment.} The $^{51}$V-NMR experiments were carried out in a field of 7.05~T in a temperature range 15\,-\,285\,K. $^{51}$V nucleus has a
spin $I\,=\,7/2$ and possesses a nuclear quadrupole moment. Therefore it is sensitive to both, the magnetic degrees of freedom and to the charge
environment. The NMR spectra were acquired by a point-by-point frequency sweeping. At each frequency point the signal has been obtained by a Fourier
transformation of the Hahn
 spin echo $\pi/2\,-\,\tau\,-\,\pi$ with the pulse separation time $\tau\,=\,13\,\mu s$. The final frequency-swept
spectrum has been calculated by sequential summing of the signals at three successive frequency points. In \figref{spectrum} a spectrum at
$T\,=\,285$\,K is shown as an example. The spectrum is broadened due to a different orientation of nanotubes in a powder sample and contains several
unresolved quadrupole satellites. The relaxation times $T_2$ and $T_1$ were measured at the frequency of the maximum intensity of the spectrum which
was assigned to the main $|\,+\,1/2\,>\,\leftrightarrow\,|\,-\,1/2\,>$ NMR transition. The transversal relaxation time $T_2$ has been determined from
the Hahn spin echo decay which cannot be described by a single exponent. However it could be very well fitted assuming the two-exponential behavior
with two distinctly different relaxation times. {To measure the longitudinal relaxation time $T_1$ a method of stimulated echo employing a pulse
sequence $\pi/2\,-\,t\,-\,\pi/2\,-\,\tau\,-\,\pi/2$ with $t\,<\,\tau$  has been used}. The time dependence of the stimulated spin echo intensity
{$A_{st}(\tau)$}, which is a measure of the decay of the longitudinal magnetization characterized by the relaxation time $T_1$, has been analyzed
following the standard equation for the spin $I\,=\,7/2$ derived in Ref.~\onlinecite{Narath67}. Also in this case the decay of {$A_{st}(\tau)$}
cannot be described by a one-center fitting. The fit requires the presence of two contributions of different weights with distinctly different
relaxation rates. Thus for the following discussion it is useful to define the ''fast'' and the ''slow'' relaxation processes. The temperature
dependence of the ''fast'' and ''slow'' transverse and longitudinal relaxation rates $T_2^{-1}$ and $T_1^{-1}$ is shown in \figref{rates}. The
''slow'' contribution to $T_2^{-1}$ is temperature independent whereas the ''fast'' one as well as both contributions to $T_1^{-1}$ increase with
increasing temperature. The ratio of the weights of these two contributions to the relaxation decay $k_{fast}\,:\,k_{slow}$ at $T\,=\,285$\,K is
close to 2.4. However the weight of the ''slow'' component strongly increases with decreasing temperature yielding
$k_{fast}\,:\,k_{slow}\,\approx\,0.7$ at $T\,=\,15$\,K (\figref{ratio}).

\textit{Discussion.} As discussed in Section~\ref{statmag} vanadium ions in VO$_x$-NTs occur in different valence states and can be magnetic and
nonmagnetic. The ions possessing an electronic magnetic moment produce a fluctuating magnetic field at the nuclei. It leads to a fast relaxation of
the nuclei of the magnetic ions due to the on-site hyperfine interaction. The magnitude of the transferred hyperfine field at the nuclei of
nonmagnetic ions is much smaller and the relaxation is more slow. The temperature independent ''slow'' component of $T_2^{-1}$ can obviously be
assigned to {the} conventional {magnetic dipolar} interaction {(see, e.g., Ref.~\onlinecite{Ansermet88})}. However with increasing temperature a more
effective relaxation channel appears. The fact that the temperature dependence of the ''fast'' $T_2^{-1}$ contribution is similar to that of the both
components of $T_1^{-1}$ (Fig.~\ref{rates}) gives evidence that the origin of this contribution is the relaxation due to the fluctuating magnetic
field {produced by electron spins at the nuclei. In this case according to the Redfield's theory \cite{Slichter89} $T_1^{-1}$ and $T_2^{-1}$ are
determined by the magnitude of the transferred field $H_i$ ($i\,=\,x,y,$ and $z$) and the correlation time $\tau_0$ of the electron spins:
\begin{eqnarray}
T_1^{-1}=\gamma_n^2[\overline{H_x^2}+\overline{H_y^2}]\tau_0/(1+\omega_n^2\tau_0^2),\\
T_2^{-1}=\gamma_n^2[\overline{H_z^2}\tau_0 + \overline{H_y^2}\tau_0/(1+\omega_n^2\tau_0^2)].
\end{eqnarray}
Here $\gamma_n$ is the nuclear gyromagnetic ratio and $\omega_n$ is the NMR frequency, respectively. In the regime of fast fluctuations,
$\omega_n\tau\,<\,1$, which is expected far away from magnetically ordered phases, the observed strong increase of the relaxation rates with
temperature (Fig.~\ref{rates}a) suggests the enhancement of $H_i$ at the positions of the nuclei owing to, e.g., an increase of the concentration of
the spin centers. Moreover,} the change of the ratio of the weight coefficients of two contributions to the relaxation decay $k_{fast}\,:\,k_{slow}$
with temperature (Fig.~\ref{ratio}) can be explained by taking into account the magnetic susceptibility data which indicate that part of magnetic
ions {form antiferromagnetic dimers} and exhibit a spin gap behavior. Usually owing to a spin gap the huge slowing down of relaxation with decreasing
temperature is observed, as, e.g. in the case of the low-dimensional vanadium oxides, it has been observed in the NMR relaxation measurements of a
two-leg spin-ladder CaV$_2$O$_5$,\cite{Iwase96,Ohama01} or a quarter-filled ladder compound NaV$_2$O$_5$.\cite{Ohama99} {The occurrence of a spin gap
$\Delta$ yields an exponential temperature dependence of the longitudinal relaxation rate $T_1^{-1}$ for $T\,<\,\Delta$: \cite{Troyer94,Naef00}
\begin{equation}
T_1^{-1}\propto \exp{(-\Delta/T)}[0.80908 - \ln{(\omega_n/T)}]. \label{spingap}
\end{equation}
With the dimer gap $\Delta\,=\,710$\, K deduced from our magnetization data one would expect a strong decrease of the nuclear relaxation below
100\,K, which is not observed experimentally (cf. \figref{rates}). Obviously,} in our case this picture is disguised by the presence of nondimerized
magnetic ions providing fast relaxation even at low temperature. These magnetic centers can be assigned to the interlayer V(3) sites as well as to
trimers in the V chains (see Section~\ref{statmag}). The nuclei in dimerized ions are slowly relaxing at low temperatures because of the non-magnetic
singlet ground state of a dimer and fast relaxing at high temperatures owing to thermally activated singlet-triplet excitations across the dimer gap.
Hence the existence of a spin gap should be seen as a change of the weight coefficients of the respective $T_1^{-1}$ contributions. Moreover, thermal
activation of dimers at high temperatures should effectively look like an increase of the concentration of magnetic centers. This should enhance the
magnitude of the fluctuating magnetic field and the relaxation rate of the ''slow-relaxing'' nuclei. Indeed, as can be seen in Fig.~\ref{ratio} with
increasing temperature the relaxation rate of the ''slow'' nuclei grows stronger than that of the ''fast'' ones. The common temperature dependence of
the ratios of the weights and of the relaxation times of two contributions (Fig.~\ref{ratio}) justifies the same origin of both dependences which can
be due to the occurrence of the spin gap for a part of magnetic vanadium ions. The weight coefficients at high temperatures provides the ratio of the
fast and slow relaxing centers approximately as 70\%\,:\,30\%. This ratio is in a fair agreement with the EELS estimation of the ratio between
magnetic V$^{4+}$ and nonmagnetic V$^{5+}$ amounting to 60\%\,:\,40\%.\cite{Liu05} Therefore in the high temperature regime it is reasonable to
associate the fast- and the slow-relaxing centers with V$^{4+}$ and V$^{5+}$ sites, respectively. From the weight ratio $k_{fast}\,:\,k_{slow}$ at
low temperatures one obtains the number of ''slow-relaxing'' nuclei amounting to $\approx\,60\,\%$ which comprises the nonmagnetic V$^{5+}$ sites as
well as the dimerized centers. As the high-temperature NMR estimate of the V$^{5+}$ sites amounts to $\approx\,30\,\%$, the concentration of the
magnetic $S\,=\,1/2$ V$^{4+}$ sites coupled in antiferromagnetic dimers is, therefore, about 30\% of all vanadium, in a good agreement with the
estimate $N_d\,=\,28$\,\% from the static magnetization data (Section~\ref{statmag}). The NMR estimate of the ''fast-relaxing'', i.e. paramagnetic
centers, at low temperatures yields $\approx\,40\,\%$ which is significantly larger than the concentration of the Curie-like spins
$N_{cw}\,=\,17$\,\% obtained from the analysis of the Curie-like susceptibility $\chi_{cw}$ (Eq.~\ref{chiCW}). This apparent contradiction can be
resolved if not only individual spins but also trimers would contribute to $\chi_{cw}$ at low temperatures, as has been suggested in the analysis of
the static magnetic data in Section~\ref{statmag}. Assuming that the nuclei in a trimer belong to the ''fast-relaxing'' centers, owing to a magnetic
ground state of a trimer, all $N_t$ sites involved in trimers will be counted in the weight of the respective contribution to the NMR relaxation rate
$T_1^{-1}$, whereas effectively only $N_t/3$ of those sites is counted in $N_{cw}$. Thus the analysis of the NMR data strongly supports the static
magnetic results, both justifying the occurrence of comparable amounts of individual spins, antiferromagnetic dimers and trimers as depicted in
\figref{Diagram}.

\section{Conclusions}

In summary, magnetization and $^{51}$V~NMR measurements of the samples of multiwalled vanadium-oxide nano\-tubes reveal a complex evolution of the
static and dynamic magnetic properties as a function of temperature. The results give strong indications that different spin arrangements beyond a
simplified picture suggested in Ref.~\onlinecite{KrusinElbaum04} are realized in VO$_x$-NTs. In particular, the data analysis  strongly suggests that
in addition to individual spins and antiferromagnetic dimers showing a spin gap of the order of 700\,K an appreciable amount of spin trimers occur in
the samples. Though the estimate of the fractional weights of different spin species summarized in \figref{Diagram} might change to some extent owing
to the experimental uncertainties, { and also because structural imperfections as well as the proportion of the majority scrolled tubes to the
minority concentric tubes may vary to some extent from sample to sample,} the data give evidence for the coexistence of different spin configurations
in comparable amounts which is related to the complex low-dimensional crystallographic structure as well as mixed valency of this nanoscale magnet.

\section{Acknowledgments}
This work was supported by the Deutsche Forschungsgemeinschaft (DFG) through project KL 1824/2.  The work of EV was supported by the DFG grant
436RUS17/38/06, joint DFG - Russian Academy of Sciences project on physics of novel materials 436RUS113/780/0-1 and also by the Russian Foundation
for Basic Research through grant No. 04-02-17137.


\references

\bibitem{Himpsel98}
{F.\ J.\ Himpsel, J.\ E.\ Ortega, G.\ J.\ Mankey, and R.\ F.\ Willis, Adv.\ Phys.\ {\bf 47} 511 (1998).}

\bibitem{Imada98}
M.\ Imada, A.\ Fujimori, and Y.\ Tokura, Rev.\ Mod.\ Phys. {\bf 70}, 1039 (1998).

\bibitem{Dagotto99}
E.\ Dagotto, Rep.\ Prog.\ Phys. {\bf 62}, 1525 (1999).

\bibitem{Tokura00}
J.\ Tokura and Y.\ Nagaosa, Science {\bf 288}, 462 (2000).

\bibitem{Orenstein00}
J.\ Orenstein and A.\ J.\ Millis, Science {\bf 288}, 468 (2000).

\bibitem{Krumeich99}
F.\ Krumeich, H.-J.\ Muhr, M.\ Niederberger, F.\ Bieri, B.\ Schnyder, and R.\  Nesper, J.\ Am.\ Chem.\ Soc. {\bf 121},
8324 (1999).

\bibitem{Worle02}
M.\ W"orle, F.\ Krumeich, F.\ Bieri, H.-J.\ Muhr, and R.\ Nesper, Z.\ Anorg.\ Allg.\ Chem. {\bf 628}, 2778 (2002).

\bibitem{Wang98}
X. Wang, L. Liu, R. Bontschev, and A. J. Jacobson, J. Chem. Soc. Chem. Commun. 1009 (1998).

\bibitem{KrusinElbaum04}
L.\ Krusin-Elbaum, M.\ N.\ Newns, H.\ Zeng, V.\ Dericke, J.\ Z.\ Sun, and R.\ Sandstrom, Nature {\bf 431}, 672 (2004).

\bibitem{Liu05}
X.\ Liu,C.\ T"aschner, A.\ Leonhardt, M.-H.\ R"ummeli, T.\ Pichler, T.\ Gemming, B.\ B"uchner, and M.\ Knupfer, Phys.\ Rev.\ B {\bf 72} 115407
(2005).

\bibitem{Gelle04}
A.\ Gelle and M.~B.\ Lepetit, Phys.\ Rev.\ Lett. {\bf 92} 236402 (2004).

\bibitem{Klingeler06}
R.\ Klingeler, B.\ B\"uchner, K.-Y.\ Choi, V.\ Kataev, U.\ Ammerahl, A.\ Revcolevschi, and J.\ Schnack, Phys.\ Rev.\ B {\bf 73}, 014426 (2006).

\bibitem{Klingeler05b}
R.\ Klingeler, N.\ Tristan, B.\ B\"uchner, M.\ H\"ucker, U.\ Ammerahl, and
A.\ Revcolevschi, Phys. Rev. B {\bf 72}, 184406 (2005).

\bibitem{Narath67}
A. Narath, Phys.\ Rev.\ {\bf 162}, 320 (1967).


\bibitem{Ansermet88}
{J.\ Ph.\ Ansermet, C.\ P.\ Slichter, and J.\ H.\ Sinflet, J.\ Chem.\ Phys. {\bf 88}, 5963 (1988). }


\bibitem{Slichter89}
{C.\ P.\ Slichter, {\it Principles of Magnetic Resonance}, 3rd ed., (Springer, New York, 1989). }

\bibitem{Iwase96}
H.\ Iwase, M.\ Isobe, Y.\ Ueda, H.\ Yasuoka, J.\ Phys.\ Soc.\ Jpn. {\bf 65}, 2397 (1996).

\bibitem{Ohama01}
T.\ Ohama, M.\ Isobe, and Y.\ Ueda, J.\ Phys.\ Soc.\ Jpn. {\bf 70}, 1801 (1996).

\bibitem{Ohama99}
T.\ Ohama, H.\ Yasuoka, M.\ Isobe, and Y.\ Ueda, Phys.\ Rev.\ B {\bf 59} 3299 (1999).


\bibitem{Troyer94}
{M.\ Troyer, H.\ Tsunetsugu, and D.\ W\"urtz, Phys.\ Rev.\ B {\bf 50} 13515 (1994). }


\bibitem{Naef00}
{F.\ Naef and X.\ Wang, Phys.\ Rev.\ Lett. {\bf 84}, 1320 (2000). }
\end{document}